\documentclass[a4paper]{article}
\usepackage{iclr2017_conference,times}
\usepackage[hidelinks,pdfpagemode=UseNone]{hyperref}  
\usepackage{graphicx}  
\usepackage{amssymb,amsmath,bm}  
\usepackage{booktabs}  
\usepackage{siunitx}  

\renewcommand{\vec}{\mathbf}
\DeclareMathOperator{\normal}{\mathcal{N}}  

\renewcommand\cite{\citep}  

\title{A Neural Parametric Singing Synthesizer}
\author{Merlijn Blaauw, Jordi Bonada
\\
\\
Music Technology Group, Universitat Pompeu Fabra, Barcelona, Spain\\
\texttt{\{merlijn.blaauw,jordi.bonada\}@upf.edu}
}

\begin{document}

\maketitle
\begin{abstract}
We present a new model for singing synthesis based on a modified version of the WaveNet architecture. Instead of modeling raw waveform, we model features produced by a parametric vocoder that separates the influence of pitch and timbre. This allows conveniently modifying pitch to match any target melody, facilitates training on more modest dataset sizes, and significantly reduces training and generation times. Our model makes frame-wise predictions using mixture density outputs rather than categorical outputs in order to reduce the required parameter count. As we found overfitting to be an issue with the relatively small datasets used in our experiments, we propose a method to regularize the model and make the autoregressive generation process more robust to prediction errors. Using a simple multi-stream architecture, harmonic, aperiodic and voiced/unvoiced components can all be predicted in a coherent manner. We compare our method to existing parametric statistical and state-of-the-art concatenative methods using quantitative metrics and a listening test. While naive implementations of the autoregressive generation algorithm tend to be inefficient, using a smart algorithm we can greatly speed up the process and obtain a system that's competitive in both speed and quality.
\end{abstract}

\section{Introduction}

Many of today's more successful singing synthesizers are based on concatenative methods. That is, they transform and concatenate short waveform units selected from an inventory of recordings of a singer. While such systems are the state-of-the-art in terms of sound quality and naturalness \cite{BonadaJ2016}, they are limited in terms of flexibility and can be difficult to extend or significantly improve upon. On the other hand, machine learning-based approaches, such as statistical parametric methods \cite{SainoK2006,OuraK2010}, are much less rigid and do allow for things such as combining data from multiple speakers, model adaptation using small amounts of training data, joint modeling of timbre and expression, etc. Unfortunately, so far these systems have been unable to match the sound quality of concatenative methods, in particular suffering from oversmoothing in frequency and time.

Recent advances in generative models for Text-to-Speech Synthesis (TTS) using Deep Neural Networks (DNNs), in particular the WaveNet model \cite{VanDenOordA2016B}, showed that model-based approaches can achieve sound quality on-par or even beyond that of concatenative systems. This model's ability to accurately generate raw speech waveform sample-by-sample, clearly shows that oversmoothing is not an issue. While directly modeling the waveform signal is very attractive, we feel that for singing voice the more traditional approach of using a parametric vocoder is better suited. Compared to speech, the melodic component of singing results in a wider range of pitch and timbre combinations, and thus a larger waveform space. We consider the required amount of training data impractical for our use case. Using a parametric vocoder effectively separates pitch and timbre, thus significantly simplifying the problem of synthesizing any melody. While a vocoder unavoidably introduces some degradation in sound quality, we consider the degradation introduced by current models still a more important factor. Thus, if we can improve the quality of the generative model, we should be able to achieve a quality closer to the upper bound the vocoder can provide, i.e. round-trip vocoder analysis-synthesis.

\section{Related work}

Our method is heavily based on a class of fully-visible probabilistic autoregressive generative models that use neural networks with similar architectures. This type of model was first proposed to model natural images (PixelCNN) \cite{VanDenOordA2016A,ReedS2016,SalimansT2017}, but was later also applied to modeling raw audio waveform (WaveNet) \cite{VanDenOordA2016B}, video (Video Pixel Networks) \cite{KalchbrennerN2016A} and text (ByteNet) \cite{KalchbrennerN2016B}.

The SampleRNN \cite{MehriS2017} model proposes an alternative architecture for unconditional raw waveform generation based on multi-scale hierarchical Recurrent Neural Networks (RNNs) rather than dilated Convolutional Neural Networks (CNNs). Other works \cite{SoteloJ2017,ArikSO2017} have extended these two architectures to include attention mechanisms to allow performing end-to-end TTS, i.e. generation conditioned on unaligned orthographic or phonetic sequences rather than aligned linguistic features.

More traditional neural parametric speech synthesizers tend to be based on feed-forward architectures such as DNNs and Mixture Density Networks (MDNs) \cite{ZenH2014}, or on recurrent architectures such as Long Short-Term Memory RNNs (LSTM-RNNs) \cite{ZenH2015}. Feed-forward networks learn a frame-wise mapping between linguistic and acoustic features, thus potentially producing discontinuous output. This is often partly mitigated by predicting static, delta and delta-delta feature distributions combined with a parameter generation algorithm that maximizes output probability \cite{TokudaK2000}. Recurrent architectures avoid this issue by propagating hidden states (and sometimes the output state) over time. In contrast, autoregressive architectures like the one we propose make predictions based on predicted past acoustic features, allowing, among other things, to better model rapid modulations such as plosive and trill consonants.

There have been several works proposing different types of singing synthesizers. The more prominent of which are based on concatenative methods \cite{BonadaJ2007,BonadaJ2016} and statistical parametric methods centered around Hidden Markov Models (HMMs) \cite{SainoK2006,OuraK2010}. While this work focuses on the generation of timbre without considering pitch or timing, an important difference between these methods is that statistical models allow joint modeling of timbre and musical expression from natural singing \cite{MaseA2010,OuraK2012}. Concatenative methods in contrast typically use disjoint modeling and specialized recordings. Many of the techniques developed for HMM-based TTS are also applicable to singing synthesis, e.g. speaker-adaptive training \cite{ShirotaK2014}. The main drawback of HMM-based approaches is that phonemes are modeled using a small number of discrete states and within each state statistics are constant. This causes excessive averaging, an overly static ``buzzy'' sound and noticeable state transitions in long sustained vowels in the case of singing. More recently, some work has also been done on using feed-forward DNNs for singing synthesis \cite{NishimuraM2016}, albeit with a somewhat limited architecture.

\section{Proposed system}

\begin{figure}[ht]
\centering
\includegraphics[width=0.8\linewidth]{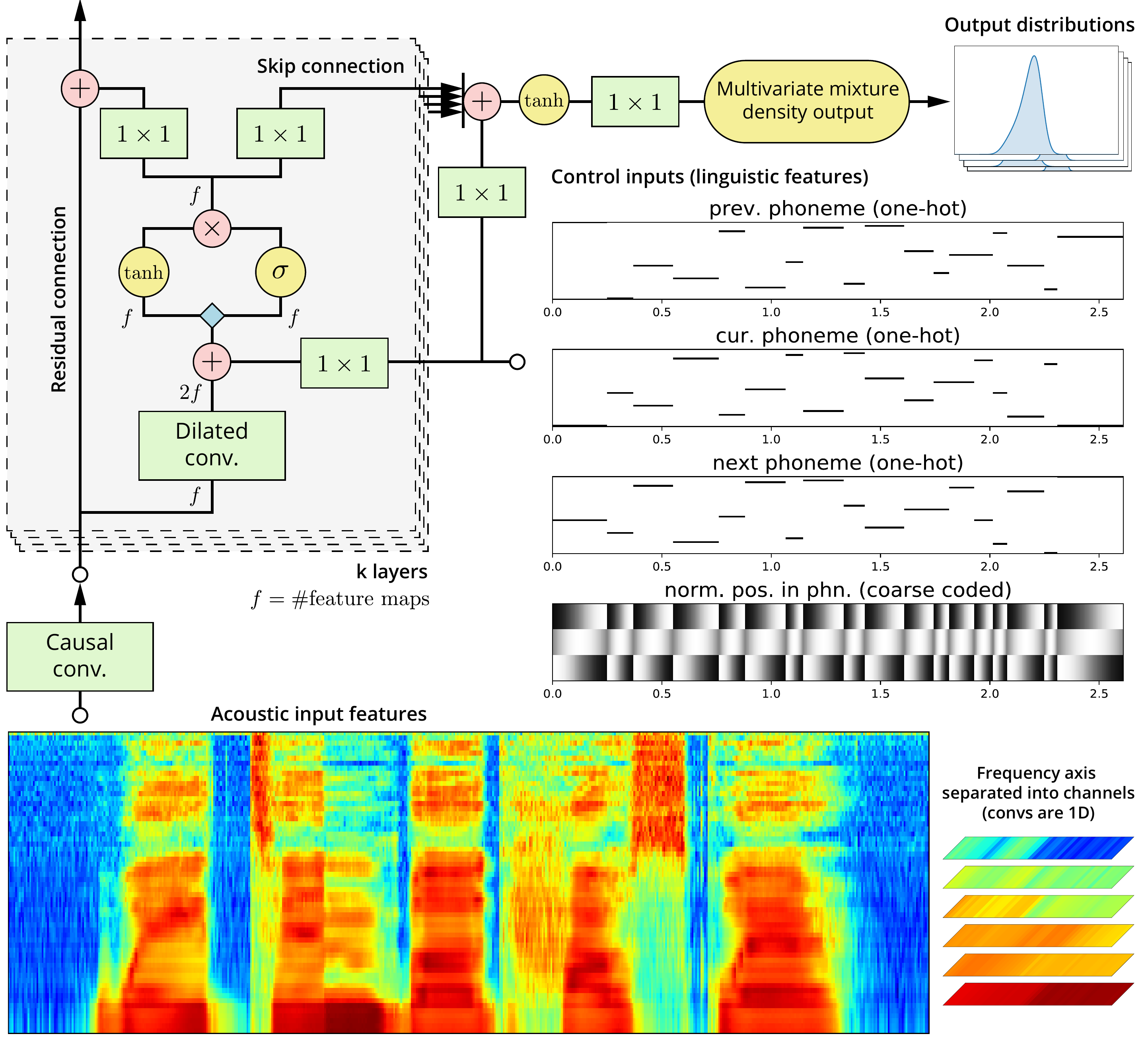}
\caption{Overview of the proposed network architecture.}
\label{fig:architecture}
\end{figure}

The architecture of our proposed model, its inputs, and its outputs are summarized in fig.~\ref{fig:architecture}. The model consists of a neural network that takes a window of past acoustic features as input and predicts a probability distribution of acoustic features corresponding to the current time step. Additionally, the network has control inputs in the form of linguistic features, which allow controlling when the model generates which phoneme.

Like its predecessors, our model is based on the idea of factorizing a joint probability as a product of conditional probabilities with some causal ordering. The conditional probability distributions are predicted by a neural network trained to maximize likelihood of a observation given past observations. To synthesize, predictions are made by sampling the distribution conditioned on past predictions, that is, in a sequential, autoregressive manner. However, while this factorization is generally done for individual variables (i.e. waveform samples or pixels), we do so for vectors of variables corresponding to a single frame,
\begin{equation}
p\left(\vec{x}_1, \dots, \vec{x}_T\right) = \prod_{t=1}^{T} p\left(\vec{x}_t \mid \vec{x}_{<t}\right),
\end{equation}
where $\vec{x}_t$ is an $N$-dimensional vector of acoustic features $\left[x_{t,1}, \dots, x_{t,N}\right]$, and $T$ is the length of the signal. In our case we consider the variables within a frame to be conditionally independent, 
\begin{equation}
p\left(\vec{x}_t \mid \vec{x}_{<t}\right) = \prod_{i=1}^{N} p\left(x_{t,i} \mid \vec{x}_{<t}\right).
\end{equation}
In other words, a single neural network predicts the parameters of a multivariate conditional distribution with diagonal covariance, corresponding to the acoustic features of a single frame.

The main reason for choosing this model is that, unlike raw audio waveform, features produced by a parametric vocoder have two dimensions, similar to (single channel) images. However, unlike images, these two dimensions are not both spatial dimensions, but rather time-frequency dimensions. The translation invariance that 2D convolutions offer is an undesirable property for the frequency (or cepstral \textit{quefrency}) dimension. Therefore, we model the features as 1D data with multiple channels. Note that these channels are only independent within the current frame; the prediction of each of the features in the current frame still depends on \emph{all} of the features of \emph{all} past frames within the receptive field (the range of input samples that affect a single output sample). This can be explained easily as all input channels of the initial causal convolution contribute to all resulting feature maps, and so on for the other convolutions.

Predicting all channels at once rather than one-by-one simplifies the models as it avoids the need for masking channels and separating them in groups. This approach is similar to \cite{SalimansT2017} where all three RGB channels of a pixel in an image are predicted at once, although in our work we do not incorporate additional linear dependencies between channel means.

The network we propose, depicted in fig.~\ref{fig:architecture}, shares most of its architecture with WaveNet. Like this model we use gated convolutional units instead of gated recurrent units such as LSTM to speed up training. The input is fed through an initial causal convolution which is then followed by stacks of $2{\times}1$ dilated convolutions \cite{YuF2015} where the dilation factor is doubled for each layer. This allows exponentially growing the model's receptive field, while linearly increasing the number of required parameters. To increase the total non-linearity of the model without excessively growing its receptive field, the dilation factor is increased up to a limit and then the sequence is repeated. We use residual and skip connections to facilitate training deeper networks \cite{HeK2016}. As we wish to control the synthesizer by inputting lyrics, we use a conditional version of the model. At every layer, before the gated non-linearity, feature maps derived from linguistic features are summed to the feature maps from the layer's main convolution. In our case we do the same thing at the output stack, similar to \cite{ReedS2016}.

\subsection{Multi-stream architecture}

Most parametric vocoders separate the speech signal into several components. In our case we obtain three \textit{feature streams}; a harmonic spectral envelope, an aperiodicity envelope and a voiced/unvoiced decision (continuous pitch is given as a control input). These components are largely independent, but their coherence is important (e.g. synthesizing a harmonic component corresponding to a voiced frame as unvoiced will generally cause artifacts). Rather than jointly modeling all data streams with a single model, we decided to model these components using independent networks. This allows us to use different architectures for each stream and, more importantly, avoids one stream possibly interfering with the other streams. For instance, the harmonic component is by far the most important, therefore we wouldn't want any other jointly modeled stream potentially reducing model capacity dedicated to this component.

To encourage predictions to be coherent, we use predictions of one network as the input of another. We currently condition the voiced/unvoiced decision on the harmonic component and the aperiodic component on both harmonic component and voiced/unvoiced decision. All the networks are similar, but can have different hyper-parameters (e.g. receptive field, early stopping). The voiced/unvoiced decision network has a Bernoulli output distribution rather than a mixture density (see~\ref{ssec:cgm}). As we found this architecture to be satisfactory, we did not investigate the many other possible variations.

\subsection{Constrained mixture density output}\label{ssec:cgm}

Many of the architectures on which we base our model predict categorical distributions, using a softmax output. The advantage of this approach is that no a priori assumptions have to be made about the (conditional) distribution of the data, allowing things such as skewed or truncated distributions, multiple modes, and so on. Drawbacks of this approach include an increase in model parameters, values are no longer ordinal, and the need to discretize data which is not naturally discrete or has high bitdepth.

Because our model predicts an entire frame at once, the issue of increased parameter count is aggravated. Instead, we opted to use a mixture density output similar to \cite{SalimansT2017}. This decision was partially motivated because in earlier versions of our model with softmax output \cite{BlaauwM2016}, we noted the predicted distributions were generally quite close to Gaussian or skewed Gaussian. In our model we use a mixture of four continuous Gaussian components, constrained in such a way that there are only four free parameters (mean, variance, skewness and a shape parameter). We found such constraints to be useful to avoid certain pathological distributions, and in our case explicitly not allowing multi-modal distributions was helpful to improve results. We also found this approach to speed up convergence compared to using categorical output.

\begin{figure}[ht]
\centering
\includegraphics[width=0.8\linewidth]{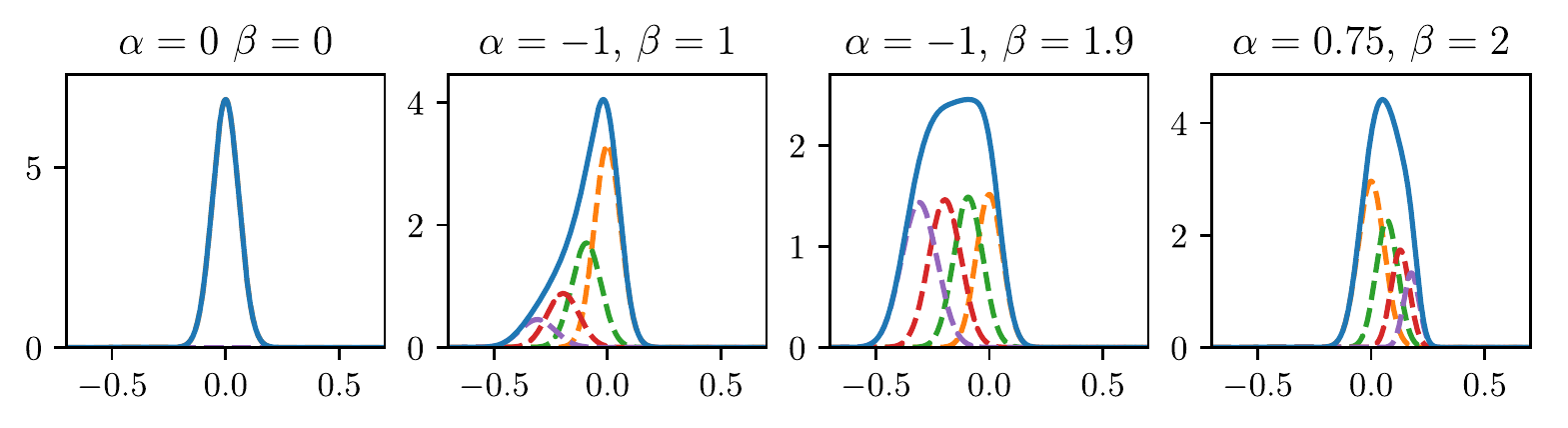}
\caption{Example distributions of the constrained mixture density output. All subplots use mean $\mu=0$ and variance $\sigma^2=\num{3.6e-3}$, but varying skewness $\alpha$ and shape $\beta$.}
\label{fig:cgm}
\end{figure}

\subsection{Robust generation by regularization}

One of the principal issues during training was that the log likelihood of the training or validation set is often not very indicative of the final synthesis quality. The most prominent symptom of this issue is that the model may occasionally produce phonemes different from those given in its control input.

One reason for this may be that the training objective does not exactly match the generation setting. During training many samples are predicted in parallel, conditioned on actual past observations. However, in generation, samples are generated one-by-one in sequential order, each one conditioned on past predictions rather than past observations. Thus the model may overfit to observations from the dataset, causing generation to fail whenever predictions diverge even slightly. We expect that this issue is more noticeable in our case because we use relatively small datasets. Additionally, the data from a parametric vocoder is inherently less structured than other types of data such as raw waveforms or natural images, that is, it tends to be smoothly varying in time, have low amounts of noise, etc.

In order to make the generation process more robust to prediction errors, we propose to use a denoising objective,
\begin{equation}
\mathcal{L} = -\log p(\vec{x}_t|\tilde{\vec{x}}_{<t}) \text{\quad with \quad} \tilde{\vec{x}}_{<t} \sim p(\tilde{\vec{x}}_{<t}|\vec{x}_{<t}),
\end{equation}
where $p(\tilde{\vec{x}}|\vec{x})$ is a Gaussian corruption distribution,
\begin{equation}
p(\tilde{\vec{x}}|\vec{x}) = \normal(\tilde{\vec{x}}; \vec{x}, \lambda I),
\end{equation}
with noise level $\lambda \geq 0$. That is, Gaussian noise is added to the input of the network, while the network is trained to predict the uncorrupted target.

When sufficiently large values of $\lambda$ are used, this technique is very effective for solving the issue of overfitting. However, the generated output can also become noticeably more noisy. One way to reduce this undesirable side effect is to apply some post processing to the predicted output distribution, much in the same vein as the \textit{temperature softmax} used in similar models, e.g. \cite{ReedS2016}. Another way to view this is that as temperature goes towards zero, the parameter generation criteria goes towards maximum likelihood, which is a commonly used in TTS systems.

We have also tried other regularization techniques, such as \textit{drop-out}, but found them to be ultimately inferior to simply injecting input noise. One possible explanation for this is that adding noise to observed context is similar to the noise introduced by prediction errors during autoregressive generation.

\subsection{Fast generation on CPU}

One drawback of autoregressive models such as the proposed system is that generation is inherently sequential and thus cannot exploit massively parallel hardware such as modern GPUs. Naive implementations of the generation algorithm thus tend to be much slower than for many other types of models. We have independently developed a fast generation algorithm using caching techniques similar to those used in \cite{RamachandranP2017,ArikSO2017}. Combining this with our frame-wise model, we can achieve speeds of 20-35$\times$ real-time on CPU. These runtimes, and the low memory and disk footprint make our system competitive with most existing systems in terms of deployability.

\section{Experimental conditions}

\subsection{Acoustic and linguistic front-end}

We use an acoustic front-end based on the WORLD vocoder \cite{MoriseM2016} with a \SI{32}{kHz} sample rate and \SI{5}{ms} hop time. The dimensionality of the harmonic component is reduced to 60 coefficients by truncated frequency warping in the cepstral domain \cite{TokudaK1994}, using an all-pole warping coefficient $\alpha=0.45$. To facilitate interpretation, the coefficients are finally converted back to frequency warped log-spectral features. The dimensionality of the aperiodic component is reduced to 4 coefficients by exploiting WORLD's inherently band-wise aperiodic analysis.

The linguistic features we use are relatively simple compared to most TTS systems as we do not model prosody. We use previous, current and next phoneme identity as one-hot encoded vectors. Additionally, we include the normalized position of the current frame within the current phoneme as a 3-state coarse coded vector, roughly corresponding to the probability of being in the beginning, middle or end of the phoneme. We do not use any features related to duration as our datasets have relatively uniform phoneme durations which may vary significantly from synthesis durations (esp. for vowels). The linguistic features are aligned to the acoustic features using a speaker-dependent Hidden Semi-Markov Model (HSMM) trained using deterministic annealing \cite{UedaN1998}.

\subsection{Datasets}

In the initial evaluation of our system, we use three voices; English male and female voices (M1, F1), and a Spanish female voice (F2). The recordings consist of short sentences which were sung at a single pitch and an approximately constant cadence. The sentences were selected to favor high diphone coverage. The Spanish dataset contains 123 sentences, while the English datasets contain 524 sentences (approx. 16 and 35 minutes respectively, including silences). Note that these datasets are small compared to the datasets typically used to train TTS systems, but this is a realistic constraint given the difficulty and cost of recording a professional singer.

\subsection{Configuration and hyper-parameters}

We use an initial causal convolution operating on 10 past values, followed by a stack of $2{\times}1$ convolutions with dilation factors [1, 2, 4, 1, 2]. This results in a total receptive field of \SI{105}{ms}. For the harmonic feature stream, we use 100 channels in the convolutions, 240 channels for the skip connections. For the aperiodicity feature stream, we use 20 channels in the convolutions, 20 channels for the skip connections. The voiced/unvoiced decision stream uses 20 channels in the convolutions and 4 channels in the skip connections. All networks use a single output stage with $\tanh$ non-linearity. We use mini-batches of 16 sequences, each with an output length of 210 frames. We use the Adam optimizer \cite{KingmaDP2015Adam} with an initial learning rate of \num{5e-4}. The model was trained for a total of 1000 epochs, taking around 8 hours on a single Titan X Pascal GPU. The entire multi-stream network contains approx. 747k trainable parameters. While we found these settings to work well experimentally, they have not been exhaustively optimized.

\section{Evaluation}

We compare our system (``NPSS'') against two alternative systems. The first is a HMM-based system (``HTS'') build using the HTS toolkit (version 2.3) \cite{ZenH2007}. Mostly standard settings were used, except for a somewhat simplified context dependency (just the two previous and two following phonemes). The second system (``IS16'') \cite{BonadaJ2016} is based on concatenative synthesis and was the highest rated system in the Interspeech 2016 Singing Synthesis Challenge.

In table~\ref{tab:quantitative_results}, we show some quantitative results comparing there systems to our proposed method. These metrics were compute over a 10\% validation split, silence frames and frames with mismatched voiced/unvoiced decision between target and prediction were excluded. The IS16 system is omitted from this table because the low redundancy of diphones in the dataset would force this system to find replacements for some diphones in the held-out validation utterances, giving the system an unfair disadvantage.

\begin{figure}[ht]
\centering
\includegraphics[width=0.8\linewidth]{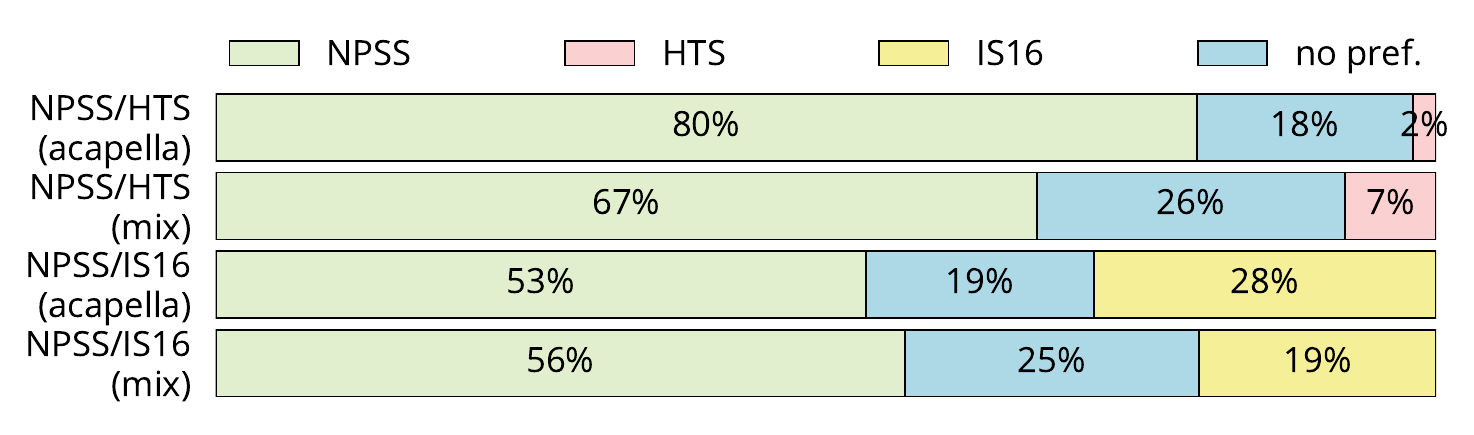}
\caption{Results of the preference test we used to qualitatively compare our model (``NPSS'') to two other approaches (``HTS'' and ``IS16''), with and without background music (acapella).}
\label{fig:pref_test}
\end{figure}

After the quantitative experiments we re-trained our models using the full datasets and synthesized one song for each of the three voices. To be able to better compare the different systems, we used pitch and phonetic timings from target recordings. From each song we extracted two short (under 10s) excerpts and made versions with and without background music. These stimuli were presented in 24 pairs to 18 participants who rated their preference between the different systems. The results of this listening tests are summarized in figure~\ref{fig:pref_test}. Full versions of the songs used in the listening test are available at:\\
\urlstyle{same}\url{http://www.dtic.upf.edu/~mblaauw/IS2017_NPSS/}

\begin{table}[th]
  \setlength{\belowcaptionskip}{1ex}
  \centering
  \caption{Quantitative results for each of the voices. The first two columns show average Mel-Cepstral Distortion (MCD) for harmonic and aperiodic components respectively. The final column shows the accuracy of the voiced/unvoiced decision prediction.}
  \label{tab:quantitative_results}
\begin{tabular}{@{}lccc@{}}
                              & \textbf{Harm. (dB)} & \textbf{Aper. (dB)} & \textbf{V/UV (acc.)} \\
\toprule
\textbf{HTS (M1)}                  & 4.24              & 0.89               & 96.86              \\
\textbf{NPSS* (M1)}                & 4.40              & 1.02               & 97.39              \\
\textbf{HTS (F1)}                  & 4.28              & 1.50               & 96.93              \\
\textbf{NPSS* (F1)}                & 4.29              & 1.62               & 97.54              \\
\textbf{HTS (F2)}                  & 4.21              & 1.27               & 98.40              \\
\textbf{NPSS* (F2)}                & 4.43              & 1.47               & 98.51              \\
\bottomrule
\end{tabular}
\end{table}

\section{Conclusions}

We presented a singing synthesizer based on neural networks that can be successfully trained on relatively small amounts of data. Listening tests showed a notable preference for our system compared to a statistical parametric system, and a moderate preference compared to a concatenative system. Interestingly, our quantitative metrics do not reflect this, showing nearly identical results. On the one hand this could be explained because the frame-wise metrics do not penalize trajectories that are overly smooth in time, yet this is perceptually important. On the other hand, these metrics do penalize small, perceptually irrelevant time-misalignments between predictions and targets, which autoregressive models may be more prone to produce. Compared to the HMM-based baseline system, we consider our method to sound noticeably less static and ``buzzy'', reproducing consonants more faithfully, and producing more natural sustained vowels. Compared to the concatenative system, we feel our method produces a similar overall sound quality. However, in certain segments, such as fast singing, subtle errors in segmentation become evident in the concatenative system. Generating phonetic contexts not in the training set is also generally handled better. The fast CPU-based autoregressive generation algorithm allows for many practical applications of our system. We hope that in the near future the flexibility offered by neural network can be explored further. In particular the area of multi-speaker training is promising, as it might help to overcome the issue of limited dataset sizes typical of singing voice.

\section{Acknowledgements}

We gratefully acknowledge the support of NVIDIA Corporation with the donation of the Titan X Pascal GPU used for this research. We also thank Zya for providing the English datasets. Voctro Labs provided the Spanish dataset and the implementation of the fast generation algorithm. This work is partially supported by the Spanish Ministry of Economy and Competitiveness under the CASAS project (TIN2015-70816-R).

\bibliographystyle{iclr2017_conference}
\bibliography{npss}

\end{document}